\documentclass[prl,twocolumn,superscriptaddress]{revtex4-1}
\usepackage{graphicx, hyperref, multirow, verbatim, fontenc}

\begin{document}

\title{Pathways Towards Ferroelectricity in Hafnia}
\author{Tran Doan Huan}
\affiliation{Department of Materials Science \& Engineering and Institute of Materials Science, University of Connecticut, 97 North Eagleville Rd., Unit 3136, Storrs, CT 06269-3136, USA}
\author{Vinit Sharma}
\affiliation{Department of Materials Science \& Engineering and Institute of Materials Science, University of Connecticut, 97 North Eagleville Rd., Unit 3136, Storrs, CT 06269-3136, USA}
\author{George A. Rossetti, Jr.}
\affiliation{Department of Materials Science \& Engineering and Institute of Materials Science, University of Connecticut, 97 North Eagleville Rd., Unit 3136, Storrs, CT 06269-3136, USA}
\author{Rampi Ramprasad}
\email{rampi@ims.uconn.edu}
\affiliation{Department of Materials Science \& Engineering and Institute of Materials Science, University of Connecticut, 97 North Eagleville Rd., Unit 3136, Storrs, CT 06269-3136, USA}

\begin{abstract}
The question of whether one can systematically identify (previously unknown) ferroelectric phases of a given material is addressed, taking hafnia (HfO$_2$) as an example. Low free energy phases at various pressures and temperatures are identified using a first-principles based structure search algorithm. Ferroelectric phases are then recognized by exploiting group theoretical principles for the symmetry-allowed displacive transitions between non-polar and polar phases. Two orthorhombic polar phases occurring in space groups $Pca2_1$ and $Pmn2_1$ are singled out as the most viable ferroelectric phases of hafnia, as they display low free energies (relative to known non-polar phases), and substantial switchable spontaneous electric polarization. These results provide an explanation for the recently observed surprising ferroelectric behavior of hafnia, and reveal pathways for stabilizing ferroelectric phases of hafnia as well as other compounds.
\end{abstract}

\pacs{XXXX}

\maketitle

Commonly known structural phases of hafnia (HfO$_2$) are centrosymmetric, and thus, non-polar. Hence, recent observations of ferroelectric behavior of hafnia thin films (when doped with Si, Zr, Y, Al or Gd) \cite{BosckeHfO2_APL2011, BosckeSiHfO2_APL2011,MuellerHfO2:12,Muller:HfO2:nanolett,Mueller12,BosckeProc11,Lomenzo:14} are rather surprising as ferroelectricity requires the presence of switchable spontaneous electrical polarization. The emergence of non-polar hafnia---as a linear high dielectric constant (or high-$\kappa$) successor to SiO$_2$---for use in modern electronic devices (e.g., field-effect transistors) is now well-established \cite{WilkHfO2,Zhu:HfO2}. If the origins of its unexpected ferroelectricity can be understood and appropriately harnessed, hafnia-based materials may find applications in nonvolatile memories and ferroelectric field effect transistors as well.

A broader question that arises within this context, and also the one that will be addressed directly in this contribution, is whether one can systematically identify ferroelectric phases of a given material system. We show that this can indeed be accomplished and ascertained, for the example of hafnia, in two steps. First, a computation-based structure search method, e.g., the minima-hopping method \cite{Goedecker:MHM,Amsler:MHM,MHM:OrganovBookChapter}, is used to identify low-energy phases at various pressures  and temperatures. Then, ferroelectric phases are singled out by applying the group theoretical symmetry reduction principles, established by Shuvalov for ferroelectricity \cite{Shuvalov}. These principles allow for the systematic identification of all possible lower symmetry proper ferroelectric phases that can result from higher-symmetry non-polar prototype (parent) phases.

\begin{figure}[b]
  \begin{center}
    \includegraphics[width= 8.25 cm]{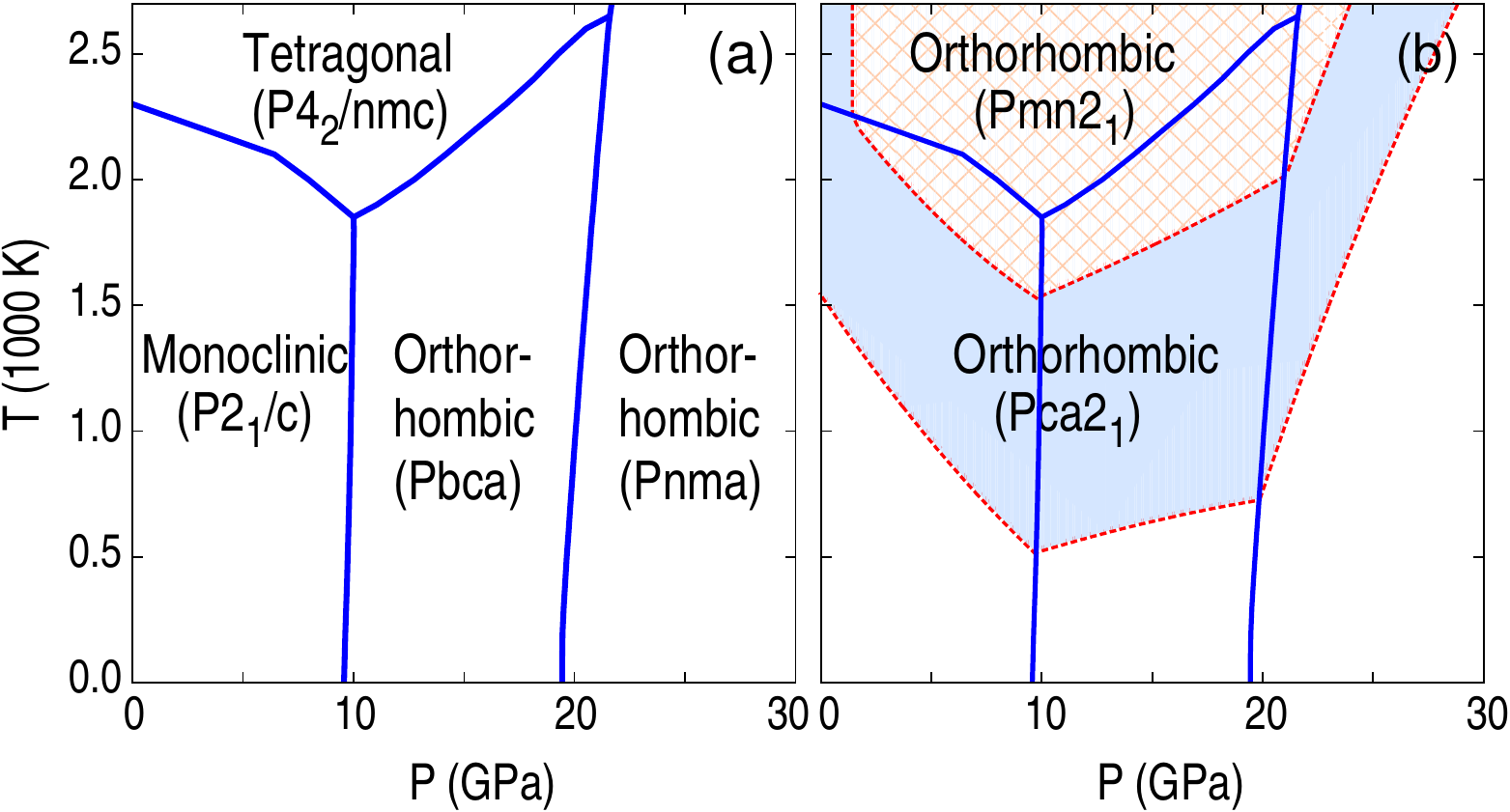}
  \caption{(Color online) (a) Computed equilibrium phase diagram of hafnia and (b) regimes in which $\Delta F(P,T)$, the free energy difference between $Pca2_1$ and $Pmn2_1$ phases and the equilibrium phases are small, i.e., $\Delta F(P,T) < k_{\rm B}T/5$.} \label{fig:phase}
  \end{center}
\end{figure}

Using this approach, we find two ferroelectric phases of hafnia, belonging to the $Pca2_1$ and $Pmn2_1$ orthorhombic space groups, which are close in free energy with the known non-polar equilibrium phases of hafnia over a wide temperature and pressure range. Figure \ref{fig:phase}(a) displays the computed equilibrium phase diagram of hafnia indicating the regimes at which the known non-polar phases are stable. This includes the low-temperature low-pressure $P2_1/c$ monoclinic phase, high-pressure $Pbca$ and $Pnma$ orthorhombic phases, and the high-temperature $P4_2/nmc$ tetragonal phase (the high-temperature $Fm\overline{3}m$ cubic aristotype is not shown). These results are consistent with available experimental data \cite{WilkHfO2,Zhu:HfO2,Ruh:HfO2,Liu:HfO2,Leger93,Ohtaka91,Ohtaka:hafnia,Tang:HfO2}. Figure \ref{fig:phase}(b) shows the temperature and pressure regimes (overlaid on the phase diagram of Figure \ref{fig:phase}(a)) when the identified ferroelectric $Pca2_1$ and $Pmn2_1$ orthorhombic phases are extremely close in free energy ($< k_{\rm B}T/5$ where $k_{\rm B}$ is the Boltzmann constant) to the equilibrium non-polar phases. The computed spontaneous polarizations of these ferroelectric phases are substantial, and switchable (at low temperatures) with small energy barriers ($\sim 10$ meV/atom) via the parent non-polar $P4_2/nmc$ tetragonal phase. Thus, $Pca2_1$ and $Pmn2_1$ are viable ferroelectric phases and may be stabilized under appropriate experimental conditions (e.g., via strain due to epitaxy or dopants).

Our density functional theory (DFT) \cite{DFT1,DFT2} calculations were performed with the {\sc abinit} package \cite{Gonze_Abinit_1, Gonze_Abinit_2}, employing the norm-conserving Hartwigsen-Goedecker-Hutter pseudopotentials \cite{HGH_Pseudo} and the Perdew-Burke-Ernzerhof exchange-correlation energies functional \cite{PBE}. Highly accurate DFT energies $E_{\rm DFT}$ were ensured by dense Monkhorst-Pack $\bf k$-point meshes \cite{monkhorst}, and a basis set of plane waves with kinetic energy up to 1100 eV. Convergence is assumed when the residual forces and stresses are smaller than $10^{-3}$~eV/\AA~ and $10^{-5}$~eV/\AA$^3$, respectively. As given in the Supplemental Material \cite{supplement}, the lattice parameters obtained by optimizing for the previously-known phases of hafnia agree very well with experimental data, implying that our computational scheme is reasonable.

Given that the ferroelectricity was observed \cite{BosckeHfO2_APL2011, BosckeSiHfO2_APL2011,MuellerHfO2:12,Muller:HfO2:nanolett,Mueller12,BosckeProc11,Lomenzo:14} in doped hafnia thin films, and that pressure is theoretically predicted \cite{Clima_PEhafnia} to play an important role in stabilizing the $Pca2_1$ polar phase, we systematically searched for low-energy structures at pressures up to 30 GPa. In the minima-hopping method~\cite{Goedecker:MHM,Amsler:MHM,MHM:OrganovBookChapter}, used for our search, the DFT energy landscape of hafnia is explored by performing consecutive short molecular-dynamics steps, travelling across local minima, progressing towards the global minimum. The initial velocities of the molecular-dynamics trajectories were chosen to lay approximately along soft mode directions, improving the efficiency of the method. Local geometry optimizations were then used to obtain the equilibrium structure. This method has been successfully applied in the past for various classes of crystalline materials, including inorganic \cite{amsler12,Huan:Alanates,Huan:Mixed} and organometallic \cite{Baldwin:SnEster1} compounds.

The structure search procedure leads to a large number of possible hafnia phases. The eleven lowest-energy cases included six non-polar and five polar phases. The six non-polar phases were the previously known $P2_1/c$ monoclinic, $P4_2/nmc$ tetragonal, $Pbca$ and $Pnma$ orthorhombic, and $Fm\overline{3}m$ cubic phases, as well as the recently predicted $P2_1/m$ monoclinic phase \cite{Zeng:HfO2}; the five polar phases were the $Pca2_1$ and $Pmn2_1$ orthorhombic, $Pm$ and $Cc$ monoclinic, and the $P1$ triclinic phases. Phonon band structure calculations, performed at zero temperature and various finite pressures up to 30 GPa using the linear response method \cite{Abinit_phonon_1, Abinit_phonon_2}, reveal that with the two exceptions of $Fm\overline{3}m$ and $P2_1/m$, the remaining structures are dynamically stable across the entire pressure range. The high-temperature $Fm\overline{3}m$ cubic structure is dynamically unstable within the whole pressure range examined, and the $P2_1/m$ structure is dynamically stable only below 15 GPa. Above this pressure, the $P2_1/m$ structure collapses to the $P4_2/nmc$ tetragonal structure (as ascertained by following the unstable phonon modes). More detailed crystallographic information, simulated XRD patterns, and phonon frequency spectra of the examined structures are given in the Supplemental Material \cite{supplement}.

\begin{figure}[t]
  \begin{center}
    \includegraphics[width= 8.25 cm]{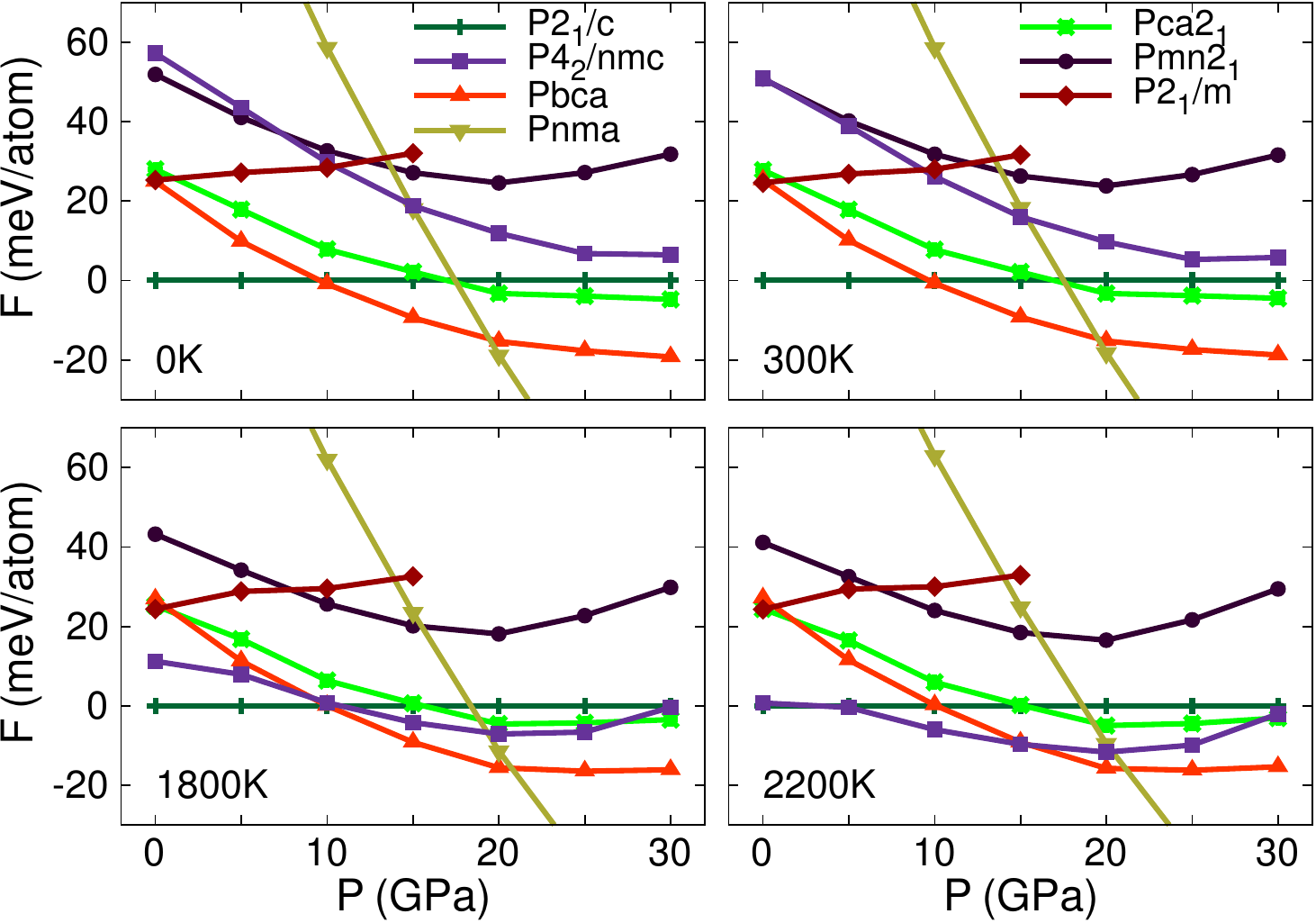}
  \caption{(Color online) Free energies $F$ calculated at $T=0$ K, $T=300$ K, $T=1800$ K, and $T=2200$ K for the identified low-energy phases of hafnia and shown as functions of pressure $P$. Data is given by symbols while curves are guides to the eye. Data corrersponding to the $Pm$, $Cc$, and $P1$ phases are not shown as they are energetically unfavorable by roughly $250$ meV/atom compared with the other phases.} \label{fig:sum}
  \end{center}
\end{figure}

From the calculated phonon band structures, free energies $F$ were computed within the harmonic approximation for the dynamically stable structures. These are shown in Fig. \ref{fig:sum} for four different temperatures across the entire pressure range considered. As established previously \cite{Liu:HfO2,Leger93,Ohtaka:hafnia,Tang:HfO2}, we found that the $P2_1/c$, $Pcba$, and $Pnma$ phases are thermodynamically stable below $10$ GPa, between $10$ GPa and $20$ GPa, and above $20$ GPa, respectively. Consistent with a recent experimental report \cite{Al-Khatatbeh:HfO2}, our additional calculations reveal that up to $100$ GPa, the $Pnma$ structure remains the most stable phase of hafnia. The stabilization of the $P4_2/nmc$ tetragonal phase is also confirmed in our work. Starting from $1800$ K and $10$ GPa, this phase is stable relative to the $P2_1/c$ monoclinic and the $Pcba$ orthorhombic phases. At ambient pressures, we predict that hafnia transforms from the $P2_1/c$ phase to the $P4_2/nmc$ phase at $T_{\rm c}\simeq 2200$ K, consistent with experiments \cite{Ohtaka:hafnia,wang_review}. The phase diagram \ref{fig:phase}(a), which was constructed from the calculated $F$, summarizes these findings.

Next, we consider the predicted polar (or non-centrosymmetric) hafnia phases, i.e., phases whose point groups do not contain an inversion operation. We note that of the five cases identified, the two orthorhombic phases $Pca2_1$ and $Pmn2_1$ are energetically competing with the equilibrium phases while the $Pm$, $Cc$ and $P1$ phases are unfavorable by roughly 250 meV/atom compared to the other phases. Figure \ref{fig:phase}(b) identifies the temperature and pressure regimes at which the $Pca2_1$ and $Pmn2_1$ phases are extremely close in free energy ($< k_BT/5$) to the corresponding equilibrium phases. These results provide a rationale for the observed ferroelectricity in hafnia under some conditions. Indeed, the $Pca2_1$ orthorhombic phase has been observed in zirconia doped with Mg, Ca, and Y \cite{Heuer:zirconia, Marshall:ZrO2,kisi,Howard:zirconia} and recently suggested \cite{Rabe:zirconia, Clima_PEhafnia} to be a low-energy polar phase of both hafnia and zirconia. The other four phases, i.e., $Pmn2_1$, $Pm$, $Cc$, and $P1$, are identified for the first time in this work.

According to Shuvalov \cite{Shuvalov}, the point symmetry $mm2$ displayed by $Pca2_1$ and $Pmn2_1$ phases, the point symmetry $m$ displayed by $Pm$ and $Cc$ phases, and the point symmetry $1$ displayed by $P1$ phase, are three ferroelectric subgroups of the $4/mmm$ tetragonal point group. In fact, the symmetry requirements placed on ferroelectric phases that can result from a parent non-polar phase through distortions are quite definitive. For instance, starting from a tetragonal prototype phase with point symmetry $4/mmm$, ferroelectric orthorhombic phases of only two types with point symmetry $mm2$ (each of them accompanied by a definite polar axis) are possible. These principles suggest that the $P4_2/nmc$ tetragonal phase of hafnia is the parent non-polar phase for all identified polar phases. Indeed, we found that all five non-centrosymmetric structures considered herein can be obtained by distorting the $P4_2/nmc$ tetragonal structure appropriately. Two most interesting cases, the $Pca2_1$ and $Pmn2_1$ orthorhombic structures can be obtained by distorting the $P4_2/nmc$ tetragonal structure along its [110] and [100] directions, respectively.

\begin{figure}[t]
  \begin{center}
    \includegraphics[width= 8.25 cm]{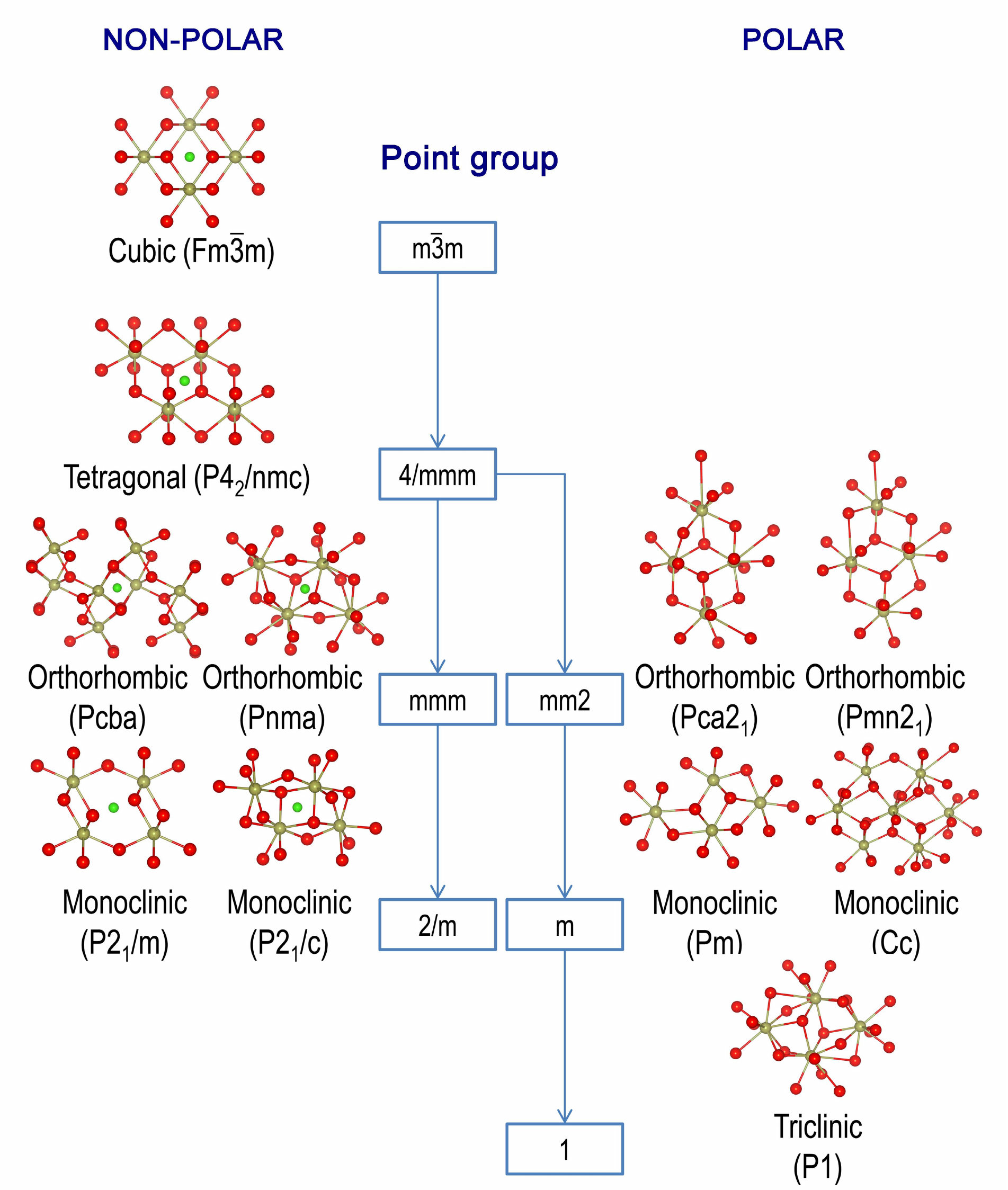}
  \caption{(Color online) Symmetry-reduction flowchart of low-energy phases of hafnia, starting from the $m\overline{3}m$ cubic point symmetry. These phases, labeled by space group symbols, are categorized into two branches, the non-polar branch ending at the $2/m$ point symmetry on the left and the polar branch leading to the $1$ point symmetry on the right. Dark yellow and red spheres represent hafnium and oxygen atoms, respectively, while green spheres locate the center of symmetry of those which are centrosymmetric.} \label{fig:struct}
  \end{center}
\end{figure}

In order to better appreciate our findings within the context of Shuvalov's symmetry-reduction principles, we show in Figure \ref{fig:struct} all the predicted low-energy phases of hafnia identified by our structure-search scheme, classified in terms of their symmetry. The right branch of Fig. 3 shows the five identified polar phases (grouped in terms of their point symmetries) emerging from the prototype $P4_2/nmc$ tetragonal phase. For completeness, we also display the non-polar phases on the left branch of Fig. \ref{fig:struct}. The highest symmetry phase at the top of the figure is the $Fm\overline{3}m$ cubic phase. Successive symmetry reduction leads to the $P4_2/nmc$ tetragonal phase, the high-pressure $Pcba$ and $Pnma$ orthorhombic phases, and finally, the ground-state $P2_1/c$ monoclinic phase and the recently-predicted $P2_1/m$ monoclinic phase \cite{Zeng:HfO2}.

We now consider just the polar $Pca2_1$ and $Pmn2_1$ orthorhombic phases, and investigate the magnitude and switchability of the polarization. Taking the $P4_2/nmc$ tetragonal phase as the reference, the spontaneous polarization was computed (through the evaluation of the Berry phase at $T=0{\rm K}$ \cite{King-Smith:polar,Resta:pol}) to be: ${\cal P}_{Pca2_1} = 52~\mu{\rm C/cm}^2$ and ${\cal P}_{Pmn2_1} = 56 ~\mu{\rm C/cm}^2$ pointing along the [001] and [100] directions, respectively (the [001] direction of the $Pca2_1$ phase corresponds to the [110] direction of the prototype $P4_2/nmc$ phase). Our calculated spontaneous polarization is in good agreement with experimentally measured values of remanent and saturation polarization (which fall within $20-50~\mu{\rm C/cm^2}$ \cite{BosckeHfO2_APL2011, BosckeSiHfO2_APL2011,MuellerHfO2:12,Muller:HfO2:nanolett,Mueller12,BosckeProc11,Lomenzo:14}) as well as with recent computations for the $Pca2_1$ phase of hafnia \cite{Rabe:zirconia,Clima_PEhafnia}.

\begin{figure*}[t]
  \begin{center}
    \includegraphics[width= 15 cm]{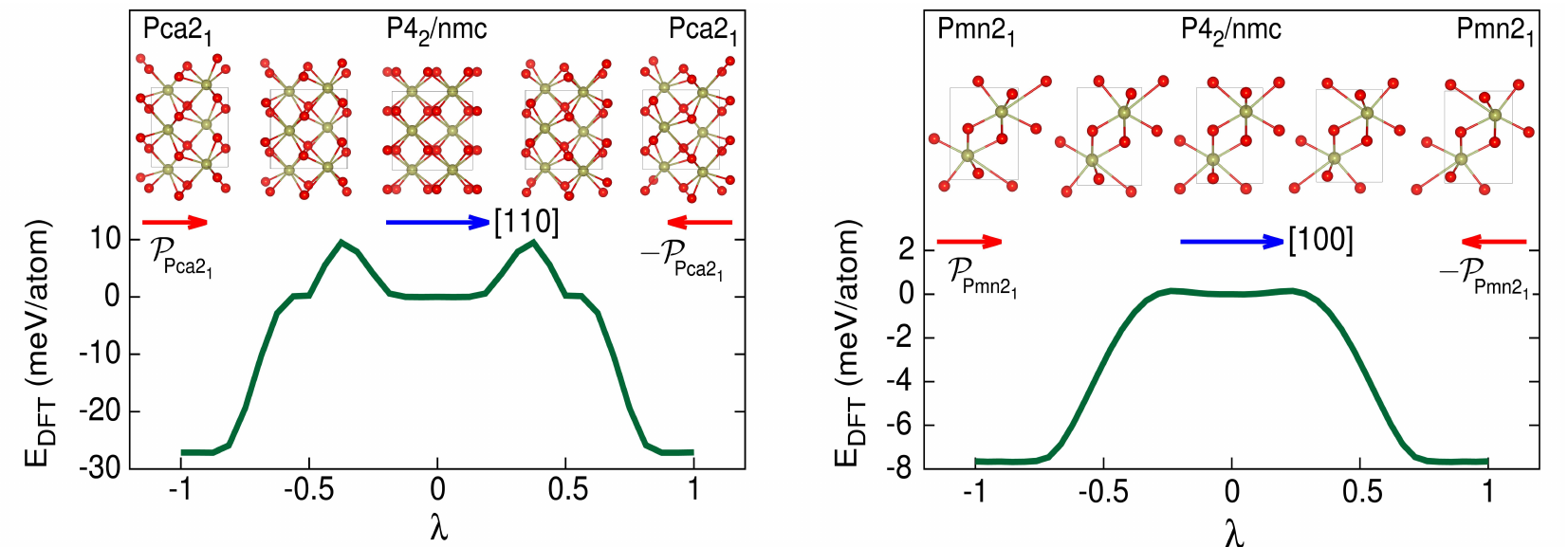}
  \caption{(Color online) Minimum (DFT) energy pathways from the ``up" state ($\lambda=-1$) to the ``down" state ($\lambda=1$) of two polar phases, $Pca2_1$ on the left panel and $Pmn2_1$ on the right panel. Illustrations for the structural arrangements along the pathways are given, using dark yellow and red spheres for hafnium and oxygen atoms, respectively. Long blue arrows indicate the deformation directions along the pathways while short red arrows represent the polarizations of the proposed polar phases. The polarization of the reference phase ($P4_2/nmc$) is zero.} \label{fig:mep}
  \end{center}
\end{figure*}

The spontaneous polarization of the $Pca2_1$ and $Pmn2_1$ phases are switchable with small energy barriers for $180$ degree switching. For each of these phases, there are two topologically equivalent variants with opposite polarization, referred to as ``up'' and ``down'' states. These variants can be obtained by deforming the prototype $P4_2/nmc$ structure along two opposite directions (see Fig. \ref{fig:mep} for an illustration). We parameterized the continuous deformation from the up to the down states by $\lambda$, a parameter scaled to range from $-1$ to $1$. The minimum (DFT) energy pathways describing the deformation leading to the $Pca2_1$ and $Pmn2_1$ phases, calculated using the generalized solid-state nudged elastic band method \cite{SSNEB}, are shown in Fig. \ref{fig:mep}. As expected, these pathways go through the $P4_2/nmc$ tetragonal phase at their centers ($\lambda=0$) before reaching the opposite polarization state. The curvatures of the calculated pathways at $\lambda=0$ are very small, indicating that although the tetragonal phase is dynamically stable, it corresponds to a very shallow minimum of the energy landscape (the corresponding phonon soft modes can be observed in the Supplemental Material \cite{supplement}). Experimentally, the transition from $P4_2/nmc$ to $Pca2_1$, hypothesized to be a polar phase of hafnia, was conjectured in Ref. \onlinecite{BosckeProc11}.

Moreover, the transitions from the $P4_2/nmc$ phase to the $Pca2_1$ and $Pmn2_1$ phases are also allowed by thermodynamic considerations. In particular, the energy barriers separating the up and the down states of the $Pca2_1$ and $Pmn2_1$ phases at $P=0$ GPa are estimated to be 40 meV/atom and 8 meV/atom, respectively. Between the $P4_2/nmc$ phase and these two ferroelectric phases, the energy barriers are smaller, i.e., 6 meV/atom for the $Pca2_1$ phase and less than 1 meV/atom for the $Pmn2_1$ phase. For zirconia, the structural phase transition from $P4_2/nmc$ to $Pca2_1$ under hydrostatic pressures has experimentally been observed \cite{kisi}.

\begin{figure}[b]
  \begin{center}
    \includegraphics[width= 7.0 cm]{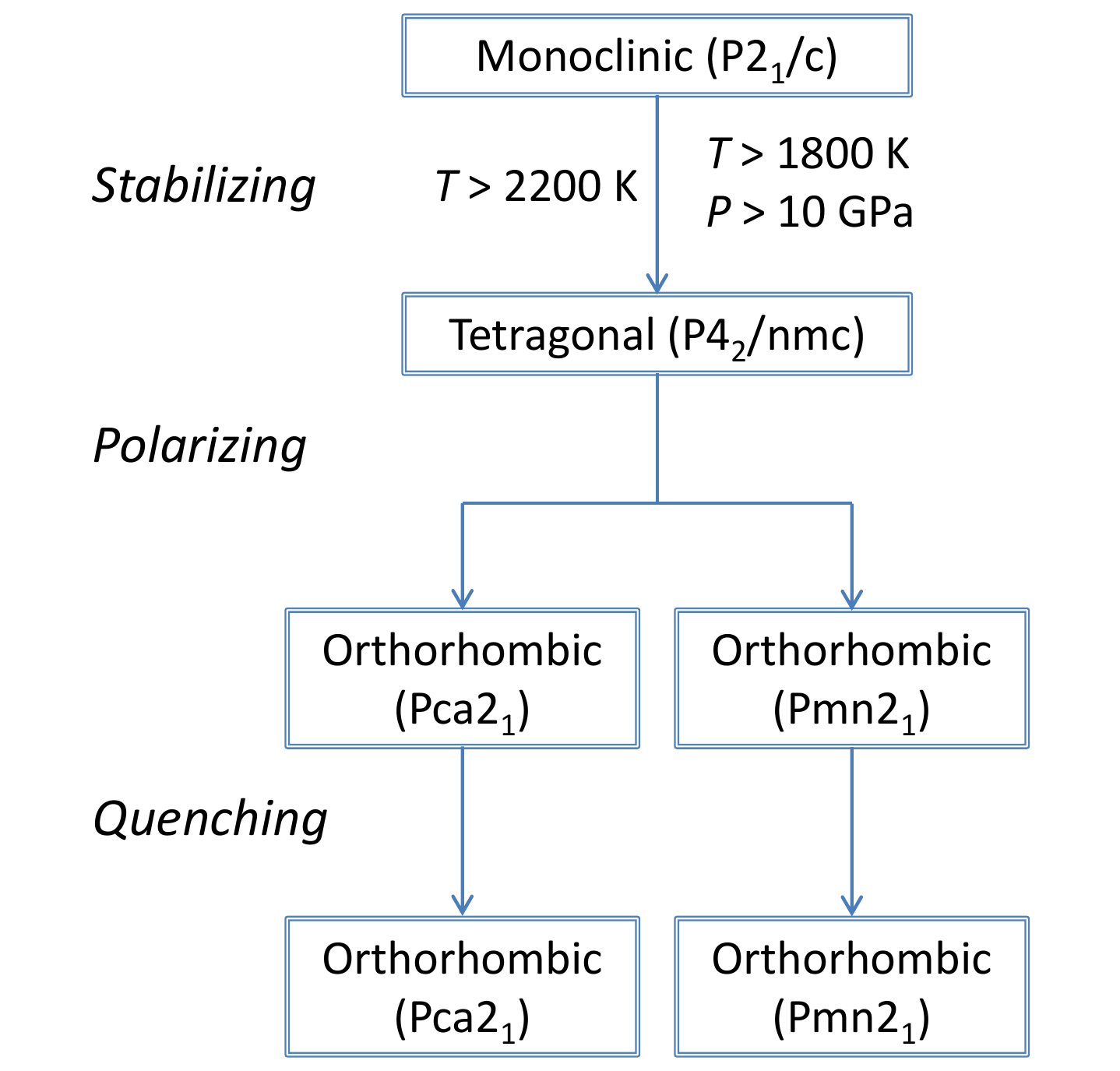}
  \caption{Possible pathways which might describe the formation process of hafnia polar phases. First, the $P4_2/nmc$ tetragonal phase is stabilized at suitable temperature and/or pressure. Next, this phase may be driven into a polar phase in the subsequent polarizing step. Finally, the polar phase may be quenched to lower temperatures and pressures.} \label{fig:paths}
  \end{center}
\end{figure}

Fig. \ref{fig:mep} suggests that while {\it en route} to the polar phases ($Pca2_1$ and $Pmn2_1$) of hafnia, the $P4_2/nmc$ tetragonal phase has to be stabilized. We note in particular that the $P4_2/nmc$ phase may be stabilized by some suitable combination of stress and/or internal/external electric field, e.g., by chemical doping or by fabrication of strain-engineered thin film structures \cite{Zhu:HfO2,WangHafnia}. In Fig. \ref{fig:paths} we tentatively sketch possible pathways which may enable the formation of polar phases even in pure hafnia. These pathways include several processes. First, the $P4_2/nmc$ tetragonal phase may be stabilized at a suitable condition, e.g., $T \simeq 2200$ K at ambient pressure or $T\simeq 1800$K at $P \simeq 10$ GPa. Next, perturbations, e.g., via an electric field, could drive the $P4_2/nmc$ phase to a ferroelectric phase, either $Pca2_1$ or $Pmn2_1$. Finally, samples may be quenched to lower temperatures and pressures, preserving the obtained polar phases.

In summary, we have systematically identified possible ferroelectric phases of hafnia from first principles by combining a low-energy structure prediction method and symmetry reduction principles established for ferroelectric phase transitions. Two of the identified orthorhombic polar phases, which belong to the $Pca2_1$ and $Pmn2_1$ space groups, are found to be extremely close in free energy to the equilibrium non-polar phases over a wide temperature and pressure window. Both polar phases can be obtained by distorting the well-known prototype non-polar $P4_2/nmc$ tetragonal structure of hafnia. The calculated spontaneous polarization is substantial and switchable (with low energy barriers), implying that the $Pca2_1$ and $Pmn2_1$ phases may provide an explanation for the recent observation of the ferroelectricity in doped hafnia films. The scheme presented here can, in principle, be used to systematically identify pathways towards ferroelectricity in other compounds as well.

The authors thank Jacob Jones for drawing their attention to this problem and for subsequent stimulating discussions. The authors also thank S. Goedecker and M. Amsler for making the minima-hopping code available.  The x-ray diffraction patterns of the examined phases were simulated by {\sc fullprof} \cite{fullprof} while their space groups were determined by {\sc findsym} \cite{findsym}. Some figures in this work were rendered with {\sc vesta} \cite{vesta}.


\begin{thebibliography}{10}

\bibitem{BosckeHfO2_APL2011}
T.~S. B{\"{o}}scke, J. M{\"{u}}ller, D. Br{\"{a}}uhaus, U. Schr{\"{o}}der, and
  U. B{\"{o}}ttger, Appl. Phys. Lett. {\bf 99},  102903  (2011).

\bibitem{BosckeSiHfO2_APL2011}
T.~S. B{\"{o}}scke, S. Teichert, D. Br{\"{a}}uhaus, J. M{\"{u}}ller, U.
  Schr{\"{o}}der, U. B{\"{o}}ttger, and T. Mikolajick, Appl. Phys. Lett. {\bf
  99},  112904  (2011).

\bibitem{MuellerHfO2:12}
S. Mueller, J. Mueller, A. Singh, S. Riedel, J. Sundqvist, U. Schroeder, and T.
  Mikolajick, Adv. Funct. Mater. {\bf 22},  2412  (2012).

\bibitem{Muller:HfO2:nanolett}
J. M{\"{u}}ller, T.~S. B{\"{o}}scke, U. Schr{\"{o}}der, S. Mueller, D.
  Br{\"{a}}uhaus, U. B{\"{o}}ttger, L. Frey, and T. Mikolajick, Nano Lett. {\bf
  12},  4318  (2012).

\bibitem{Mueller12}
S. Mueller, C. Adelmann, A. Singh, A. Van~Elshocht, U. Schroeder, and T.
  Mikolajick, ECS J. Solid State Sci. Technol. {\bf 1},  N123  (2012).

\bibitem{BosckeProc11}
T.~S. B{\"{o}}scke, J. M{\"{u}}ller, D. Br{\"{a}}uhaus, U. Schr{\"{o}}der, and
  U. B{\"{o}}ttger,  in {\em Proceedings of the 2011 IEEE International
  Electron Devices Meeting (IEDM)} (IEEE, Washington, D.C., 2011), pp.\
  24.5.1--24.5.4.

\bibitem{Lomenzo:14}
P.~D. Lomenzo, P. Zhao, Q. Takmeel, S. Moghaddam, T. Nishida, M. Nelson, C.~M.
  Fancher, E.~D. Grimley, X. Sang, J.~M. LeBeau, and J.~L. Jones, J. Vac. Sci.
  Technol. B {\bf 32},  03D123  (2014).

\bibitem{WilkHfO2}
G.~D. Wilk, R.~M. Wallace, and J.~M. Anthony, J. Appl. Phys. {\bf 87},  484
  (2000).

\bibitem{Zhu:HfO2}
H. Zhu, C. Tang, L. Fonseca, and R. Ramprasad, J. Mater. Sci. {\bf 47},  7399
  (2012).

\bibitem{Goedecker:MHM}
S. Goedecker, J. Chem. Phys. {\bf 120},  9911  (2004).

\bibitem{Amsler:MHM}
M. Amsler and S. Goedecker, J. Chem. Phys. {\bf 133},  224104  (2010).

\bibitem{MHM:OrganovBookChapter}
S. Goedecker,  in {\em Modern Methods of Crystal Structure Prediction}, edited
  by A.~R. Oganov (Wiley-VCH, Weinheim, 2011), Chap.~7, pp.\ 147--180.

\bibitem{Shuvalov}
L.~A. Shuvalov, J. Phys. Soc. Jpn. {\bf 28},  38  (1970).

\bibitem{Ruh:HfO2}
R. Ruh and P.~W.~R. Corfield, J. Am. Ceram. Soc. {\bf 53},  126  (1970).

\bibitem{Liu:HfO2}
L.-G. Liu, J. Phys. Chem. Solids {\bf 41},  331  (1980).

\bibitem{Leger93}
J.~M. Leger, A. Atouf, P.~E. Tomaszewski, and A.~S. Pereira, Phys. Rev. B {\bf
  48},  93  (1993).

\bibitem{Ohtaka91}
O. Ohtaka, T. Yamanaka, and S. Kume, Nippon Seramikkusu Kyokai Gakujutsu
  Ronbunshi {\bf 99},  826  (1991), (J. Ceram. Soc. Jpn.).

\bibitem{Ohtaka:hafnia}
O. Ohtaka, H. Fukui, T. Kunisada, T. Fujisawa, K. Funakoshi, W. Utsumi, T.
  Irifune, K. Kuroda, and T. Kikegawa, J. Am. Ceram. Soc. {\bf 84},  1369
  (2001).

\bibitem{Tang:HfO2}
J. Tang, M. Kai, Y. Kobayashi, S. Endo, O. Shimomura, T. Kikegawa, and T.
  Ashida,  in {\em Properties of Earth and Planetary Materials at High Pressure
  and Temperature}, edited by M.~H. Manghnani and T. Yagi (American Geophysical
  Union, Washington, D. C., 2013), pp.\ 401--407.

\bibitem{DFT1}
P. Hohenberg and W. Kohn, Phys. Rev. {\bf 136},  B864  (1964).

\bibitem{DFT2}
W. Kohn and L. Sham, Phys. Rev. {\bf 140},  A1133  (1965).

\bibitem{Gonze_Abinit_1}
X. Gonze, B. Amadon, P.-M. Anglade, J.-M. Beuken, F. Bottin, P. Boulanger, F.
  Bruneval, D. Caliste, R. Caracas, M. C{\^{o}}t{\'{e}}, T. Deutsch, L.
  Genovese, P. Ghosez, M. Giantomassi, S. Goedecker, D. Hamann, P. Hermet, F.
  Jollet, G. Jomard, S. Leroux, M. Mancini, S. Mazevet, M. Oliveira, G. Onida,
  Y. Pouillon, T. Rangel, G.-M. Rignanese, D. Sangalli, R. Shaltaf, M. Torrent,
  M. Verstraete, G. Zerah, and J. Zwanziger, Comput. Phys. Commun. {\bf 180},
  2582  (2009).

\bibitem{Gonze_Abinit_2}
X. Gonze, G.~M. Rignanese, M. Verstraete, J.-M. Beuken, Y. Pouillon, R.
  Caracas, F. Jollet, M. Torrent, G. Zerah, M. Mikami, P. Ghosez, M. Veithen,
  J.-Y. Raty, V. Olevano, F. Bruneval, L. Reining, R. Godby, G. Onida, D.~R.
  Hamann, and D.~C. Allan, Zeit. Kristallogr. {\bf 220},  558  (2005).

\bibitem{HGH_Pseudo}
C. Hartwigsen, S. Goedecker, and J. Hutter, Phys. Rev. B {\bf 58},  3641
  (1998).

\bibitem{PBE}
J.~P. Perdew, K. Burke, and M. Ernzerhof, Phys. Rev. Lett. {\bf 77},  3865
  (1996).

\bibitem{monkhorst}
H.~J. Monkhorst and J.~D. Pack, Phys. Rev. B {\bf 13},  5188  (1976).

\bibitem{supplement}
See Supplemental Material for more information reported in this paper.

\bibitem{Clima_PEhafnia}
S. Clima, D.~J. Wouters, C. Adelmann, T. Schenk, U. Schroeder, M. Jurczak, and
  G. Pourtois, Appl. Phys. Lett. {\bf 104},  092906  (2014).

\bibitem{amsler12}
M. Amsler, J.~A. Flores-Livas, T.~D. Huan, S. Botti, M.~A.~L. Marques, and S.
  Goedecker, Phys. Rev. Lett. {\bf 108},  205505  (2012).

\bibitem{Huan:Alanates}
T.~D. Huan, M. Amsler, M.~A.~L. Marques, S. Botti, A. Willand, and S.
  Goedecker, Phys. Rev. Lett. {\bf 110},  135502  (2013).

\bibitem{Huan:Mixed}
T.~D. Huan, M. Amsler, R. Sabatini, V.~N. Tuoc, N.~B. Le, L.~M. Woods, N.
  Marzari, and S. Goedecker, Phys. Rev. B {\bf 88},  024108  (2013).

\bibitem{Baldwin:SnEster1}
A.~F. Baldwin, R. Ma, A. Kumar~M.K., T.~D. Huan, C. Wang, J.~E. Marszalek, M.
  Cakmak, R. Ramprasad, and G.~A. Sotzing (unpublished).

\bibitem{Zeng:HfO2}
Q. Zeng, A.~R. Oganov, A.~O. Lyakhov, C. Xie, X. Zhang, J. Zhang, Q. Zhu, B.
  Wei, I. Grigorenko, L. Zhang, and L. Cheng, Acta Crystallogr. Sect. C {\bf
  70},  76  (2014).

\bibitem{Abinit_phonon_1}
C. Lee and X. Gonze, Phys. Rev. B {\bf 51},  8610  (1995).

\bibitem{Abinit_phonon_2}
X. Gonze and C. Lee, Phys. Rev. B {\bf 55},  10355  (1997).

\bibitem{Al-Khatatbeh:HfO2}
Y. Al-Khatatbeh, K.~K.~M. Lee, and B. Kiefer, Phys. Rev. B {\bf 82},  144106
  (2010).

\bibitem{wang_review}
J. Wang, H. Li, and R. Stevens, J. Mater. Sci. {\bf 27},  5397  (1992).

\bibitem{Heuer:zirconia}
A. Heuer, V. Lanteri, S. Farmer, R. Chaim, R. Lee, B. Kibbel, and R. Dickerson,
  J. Mater. Sci. {\bf 24},  124  (1989).

\bibitem{Marshall:ZrO2}
D.~B. Marshall, M.~R. Jarnes, and J.~R. Porter, J. Am. Ceram. Soc. {\bf 72},
  218  (1989).

\bibitem{kisi}
E.~H. Kisi, J. Am. Ceram. Soc. {\bf 81},  741  (1998).

\bibitem{Howard:zirconia}
C.~J. Howard, E.~H. Kisi, R.~B. Roberts, and R.~J. Hill, J. Am. Ceram. Soc.
  {\bf 73},  2828  (1990).

\bibitem{Rabe:zirconia}
S.~E. Reyes-Lillo, K.~F. Garrity, and K.~M. Rabe, arXiv:1403.3878
  (unpublished).

\bibitem{King-Smith:polar}
R.~D. King-Smith and D. Vanderbilt, Phys. Rev. B {\bf 47},  1651  (1993).

\bibitem{Resta:pol}
R. Resta, Rev. Mod. Phys. {\bf 66},  899  (1994).

\bibitem{SSNEB}
D. Sheppard, P. Xiao, W. Chemelewski, D.~D. Johnson, and G. Henkelman, J. Chem.
  Phys. {\bf 136},  074103  (2012).

\bibitem{WangHafnia}
L.~G. Wang, Y. Xiong, W. Xiao, L. Cheng, J. Du, H. Tu, and A. van~de Walle,
  Appl. Phys. Lett. {\bf 104},  201903,  (2014).

\bibitem{fullprof}
J. Rodr\'{i}guez-Carvajal, Physica B {\bf 192},  55  (1993).

\bibitem{findsym}
FINDSYM,
  \href{http://stokes.byu.edu/findsym.html}{http://stokes.byu.edu/findsym.html}.

\bibitem{vesta}
K. Momma and F. Izumi, J. Appl. Crystallogr. {\bf 41},  653  (2008).

\end{thebibliography}

\end{document}


\title{Supplemental Material: Pathways Towards Ferroelectricity in Hafnia}

\author{Tran Doan Huan}
\affiliation{Department of Materials Science \& Engineering and Institute of Materials Science, University of Connecticut, 97 North Eagleville Rd., Unit 3136, Storrs, CT 06269-3136, USA}
\author{Vinit Sharma}
\affiliation{Department of Materials Science \& Engineering and Institute of Materials Science, University of Connecticut, 97 North Eagleville Rd., Unit 3136, Storrs, CT 06269-3136, USA}
\author{George A. Rossetti, Jr.}
\affiliation{Department of Materials Science \& Engineering and Institute of Materials Science, University of Connecticut, 97 North Eagleville Rd., Unit 3136, Storrs, CT 06269-3136, USA}
\author{Rampi Ramprasad}
\email{rampi@ims.uconn.edu}
\affiliation{Department of Materials Science \& Engineering and Institute of Materials Science, University of Connecticut, 97 North Eagleville Rd., Unit 3136, Storrs, CT 06269-3136, USA}

\maketitle


\begin{longtable}{l r r r r r r}
\caption{Crystallographic information of the low-energy structures of hafnia discovered in this work. For each structure, cell parameters are given while for each atom, Wyckoff possition and coordinates ($x$, $y$, and $z$) are given. Data are calculated at $P=0$GPa.}\\ 
\endfirsthead
\multicolumn{7}{c}%
{{\bfseries \tablename\ \thetable{} -- continued from previous page}} \\
\hline

\endhead

\hline
\multicolumn{7}{r}{{to be continued .. }} \\
\hline
\endfoot

\hline \hline
\endlastfoot

\hline
\hline
$Pmn2_1$ & $a$ (\AA) & $b$ (\AA) & $c$ (\AA) & $\alpha$ ($^\circ$) & $\beta$ ($^\circ$) & $\gamma$ ($^\circ$)\\
\cline{2-7}
(31) & $3.415$ & $5.182$ & $3.834$ & $90$ & $90$ & $90$ \\
\hline
Atom &  \multicolumn{2}{r}{$x$} & \multicolumn{2}{r}{$y$} & \multicolumn{2}{r}{$z$} \\
\hline
 Hf (2a) &\multicolumn{2}{r}{$ 0.00000 $}&\multicolumn{2}{r}{ $-0.26306 $}&\multicolumn{2}{r}{$-0.24442$}\\  
 O  (2a) &\multicolumn{2}{r}{$ 0.00000 $}&\multicolumn{2}{r}{ $ 0.07198 $}&\multicolumn{2}{r}{$ 0.41230$}\\
 O  (2a) &\multicolumn{2}{r}{$ 0.00000 $}&\multicolumn{2}{r}{ $-0.44299 $}&\multicolumn{2}{r}{$ 0.26207$}\\
\hline
\hline
$Pm$ & $a$ (\AA) & $b$ (\AA) & $c$ (\AA) & $\alpha$ ($^\circ$) & $\beta$ ($^\circ$) & $\gamma$ ($^\circ$)\\
\cline{2-7}
(6)  & $5.364$ & $3.072$ & $6.674$ & $90$ & $84.035$ & $90$ \\
\hline
Atom &  \multicolumn{2}{r}{$x$} & \multicolumn{2}{r}{$y$} & \multicolumn{2}{r}{$z$} \\
\hline
 Hf (1a) &\multicolumn{2}{r}{$  0.36880 $}&\multicolumn{2}{r}{ $   0.00000 $}&\multicolumn{2}{r}{ $  -0.44219$}\\           
 Hf (1b) &\multicolumn{2}{r}{$ -0.11913 $}&\multicolumn{2}{r}{ $   0.50000 $}&\multicolumn{2}{r}{ $   0.36370$}\\           
 Hf (1b) &\multicolumn{2}{r}{$  0.32687 $}&\multicolumn{2}{r}{ $   0.50000 $}&\multicolumn{2}{r}{ $   0.04966$}\\           
 Hf (1b) &\multicolumn{2}{r}{$ -0.15689 $}&\multicolumn{2}{r}{ $   0.50000 $}&\multicolumn{2}{r}{ $  -0.18305$}\\           
 O  (1b) &\multicolumn{2}{r}{$  0.48014 $}&\multicolumn{2}{r}{ $   0.50000 $}&\multicolumn{2}{r}{ $  -0.25367$}\\           
 O  (1a) &\multicolumn{2}{r}{$ -0.23142 $}&\multicolumn{2}{r}{ $   0.00000 $}&\multicolumn{2}{r}{ $  -0.43264$}\\           
 O  (1a) &\multicolumn{2}{r}{$  0.16687 $}&\multicolumn{2}{r}{ $   0.00000 $}&\multicolumn{2}{r}{ $   0.27744$}\\           
 O  (1b) &\multicolumn{2}{r}{$ -0.06142 $}&\multicolumn{2}{r}{ $   0.50000 $}&\multicolumn{2}{r}{ $   0.08486$}\\           
 O  (1b) &\multicolumn{2}{r}{$  0.11456 $}&\multicolumn{2}{r}{ $   0.50000 $}&\multicolumn{2}{r}{ $  -0.42294$}\\           
 O  (1b) &\multicolumn{2}{r}{$ -0.48718 $}&\multicolumn{2}{r}{ $   0.50000 $}&\multicolumn{2}{r}{ $   0.34220$}\\           
 O  (1a) &\multicolumn{2}{r}{$  0.12953 $}&\multicolumn{2}{r}{ $   0.00000 $}&\multicolumn{2}{r}{ $  -0.14779$}\\           
 O  (1a) &\multicolumn{2}{r}{$ -0.40497 $}&\multicolumn{2}{r}{ $   0.00000 $}&\multicolumn{2}{r}{ $   0.00050$}\\           
\hline
\hline
$Cc$ & $a$ (\AA) & $b$ (\AA) & $c$ (\AA) & $\alpha$ ($^\circ$) & $\beta$ ($^\circ$) & $\gamma$ ($^\circ$)\\
\cline{2-7}
(9)  & $7.282$ & $7.317$ & $6.104$ & $90$ & $122.891$ & $90$ \\
\hline
Atom &  \multicolumn{2}{r}{$x$} & \multicolumn{2}{r}{$y$} & \multicolumn{2}{r}{$z$} \\
\hline
 Hf (4a) &\multicolumn{2}{r}{$  -0.15563  $}&\multicolumn{2}{r}{ $  0.14828 $}&\multicolumn{2}{r}{ $   0.40261$}\\
 Hf (4a) &\multicolumn{2}{r}{$   0.33641  $}&\multicolumn{2}{r}{ $  0.13000 $}&\multicolumn{2}{r}{ $   0.39879$}\\
 O  (4a) &\multicolumn{2}{r}{$  -0.48544  $}&\multicolumn{2}{r}{ $  0.29636 $}&\multicolumn{2}{r}{ $  -0.28442$}\\
 O  (4a) &\multicolumn{2}{r}{$  -0.30707  $}&\multicolumn{2}{r}{ $  0.38646 $}&\multicolumn{2}{r}{ $   0.18080$}\\
 O  (4a) &\multicolumn{2}{r}{$   0.48426  $}&\multicolumn{2}{r}{ $  0.09847 $}&\multicolumn{2}{r}{ $   0.17835$}\\
 O  (4a) &\multicolumn{2}{r}{$  -0.18951  $}&\multicolumn{2}{r}{ $  0.10203 $}&\multicolumn{2}{r}{ $  -0.28632$}\\
\hline
\hline
$P1$ & $a$ (\AA) & $b$ (\AA) & $c$ (\AA) & $\alpha$ ($^\circ$) & $\beta$ ($^\circ$) & $\gamma$ ($^\circ$)\\
\cline{2-7}
(1)  & $3.244$ & $5.674$ & $6.832$ & $91.24$ & $103.73$ & $89.99$ \\
\hline
Atom &  \multicolumn{2}{r}{$x$} & \multicolumn{2}{r}{$y$} & \multicolumn{2}{r}{$z$} \\
\hline
 Hf  (1a) &\multicolumn{2}{r}{$ 0.561145 $}&\multicolumn{2}{r}{ $0.719476 $}&\multicolumn{2}{r}{ $0.122273$}\\     
 Hf  (1a) &\multicolumn{2}{r}{$ 0.438855 $}&\multicolumn{2}{r}{ $0.280524 $}&\multicolumn{2}{r}{ $0.877727$}\\     
 Hf  (1a) &\multicolumn{2}{r}{$ 0.807632 $}&\multicolumn{2}{r}{ $0.758522 $}&\multicolumn{2}{r}{ $0.615329$}\\     
 Hf  (1a) &\multicolumn{2}{r}{$ 0.192368 $}&\multicolumn{2}{r}{ $0.241478 $}&\multicolumn{2}{r}{ $0.384671$}\\     
 O   (1a) &\multicolumn{2}{r}{$ 0.603608 $}&\multicolumn{2}{r}{ $0.345585 $}&\multicolumn{2}{r}{ $0.207160$}\\     
 O   (1a) &\multicolumn{2}{r}{$ 0.396392 $}&\multicolumn{2}{r}{ $0.654415 $}&\multicolumn{2}{r}{ $0.792840$}\\     
 O   (1a) &\multicolumn{2}{r}{$ 0.808026 $}&\multicolumn{2}{r}{ $0.366125 $}&\multicolumn{2}{r}{ $0.616133$}\\     
 O   (1a) &\multicolumn{2}{r}{$ 0.191974 $}&\multicolumn{2}{r}{ $0.633875 $}&\multicolumn{2}{r}{ $0.383867$}\\     
 O   (1a) &\multicolumn{2}{r}{$ 0.676170 $}&\multicolumn{2}{r}{ $0.974592 $}&\multicolumn{2}{r}{ $0.352351$}\\     
 O   (1a) &\multicolumn{2}{r}{$ 0.323830 $}&\multicolumn{2}{r}{ $0.025408 $}&\multicolumn{2}{r}{ $0.647649$}\\     
 O   (1a) &\multicolumn{2}{r}{$ 1.000000 $}&\multicolumn{2}{r}{ $0.000000 $}&\multicolumn{2}{r}{ $1.000000$}\\     
 O   (1a) &\multicolumn{2}{r}{$ 0.000000 $}&\multicolumn{2}{r}{ $0.500000 $}&\multicolumn{2}{r}{ $1.000000$}\\     
\end{longtable}

\begin{table}[H]
\begin{center}
\caption{A short summary of the lowest-energy structures of hafnia. References (as shown in the main text) are given for data originated from other works while dagger symbols ($\dagger$) indicate our calculated data.}\label{table:sum}
\begin{tabular}{cccccccc}
\hline
\hline
Structure  & $P$(GPa) &$a$(\AA) & $b$(\AA) & $c$(\AA) & $\beta(^\circ)$ & Ref. & ${\cal P}\left(\frac{\mu{\rm C}}{cm^2}\right)$\\
\hline
$P2_1/c$   & $0$ & $5.14$ & $5.20$ & $5.31$& $99.8$ &$\dagger$& $0$\\
           & $0$ & $5.12$ & $5.17$ & $5.29$& $99.2$ & [14]& $-$  \\
\hline
$Pbca$ & $0$ & $5.08$ & $10.03$ & $5.27$ & $90$ & $\dagger$& $0$ \\
       & $0$ & $5.06$ & $10.02$ & $5.23$ & $90$ & [17]  & $-$\\
\hline
$Pnma$ & $0$ & $5.54$ & $3.31$ & $6.55$ & $90$ &$\dagger$& $0$ \\
       & $0$ & $5.57$ & $3.30$ & $6.45$ & $90$ & [19]    & $-$\\
       & $38.5$ & $5.37$ & $3.17$ & $6.29$ & $90$ &$\dagger$& $-$ \\
       & $38.5$ & $5.41$ & $3.18$ & $6.28$ & $90$ & [19] & $-$\\
\hline
$P4_2/nmc$ & $0$ & $3.58$ & $3.58$ & $5.28$ & $90$ &$\dagger$& $0$ \\
           & $0$ & $3.58$ & $3.58$ & $5.20$ & $90$ & [33] & $-$ \\
\hline
$Pca2_1$ & $0$ & $5.29$ & $5.01$ & $5.08$ & $90$ &$\dagger$& $52$ \\
         & $0$ & $5.10$ & $4.90$ & $4.92$ & $90$ &[33] & $-$ \\
\hline
$P2_1/m$ & $0$ & $3.87$ & $3.46$ & $5.43$ & $104.7$ &$\dagger$& $0$ \\
         & $0$ & $3.82$ & $3.43$ & $5.38$ & $104.5$ &[33]& $-$ \\
\hline
$Pmn2_1$ & $0$ & $3.41$ & $5.18$ & $3.83$ & $90$ &$\dagger$& $56$ \\
\hline
\hline
\end{tabular}
\end{center}
\end{table}

\begin{figure*}[t]
 \begin{center}
\includegraphics[width=13cm]{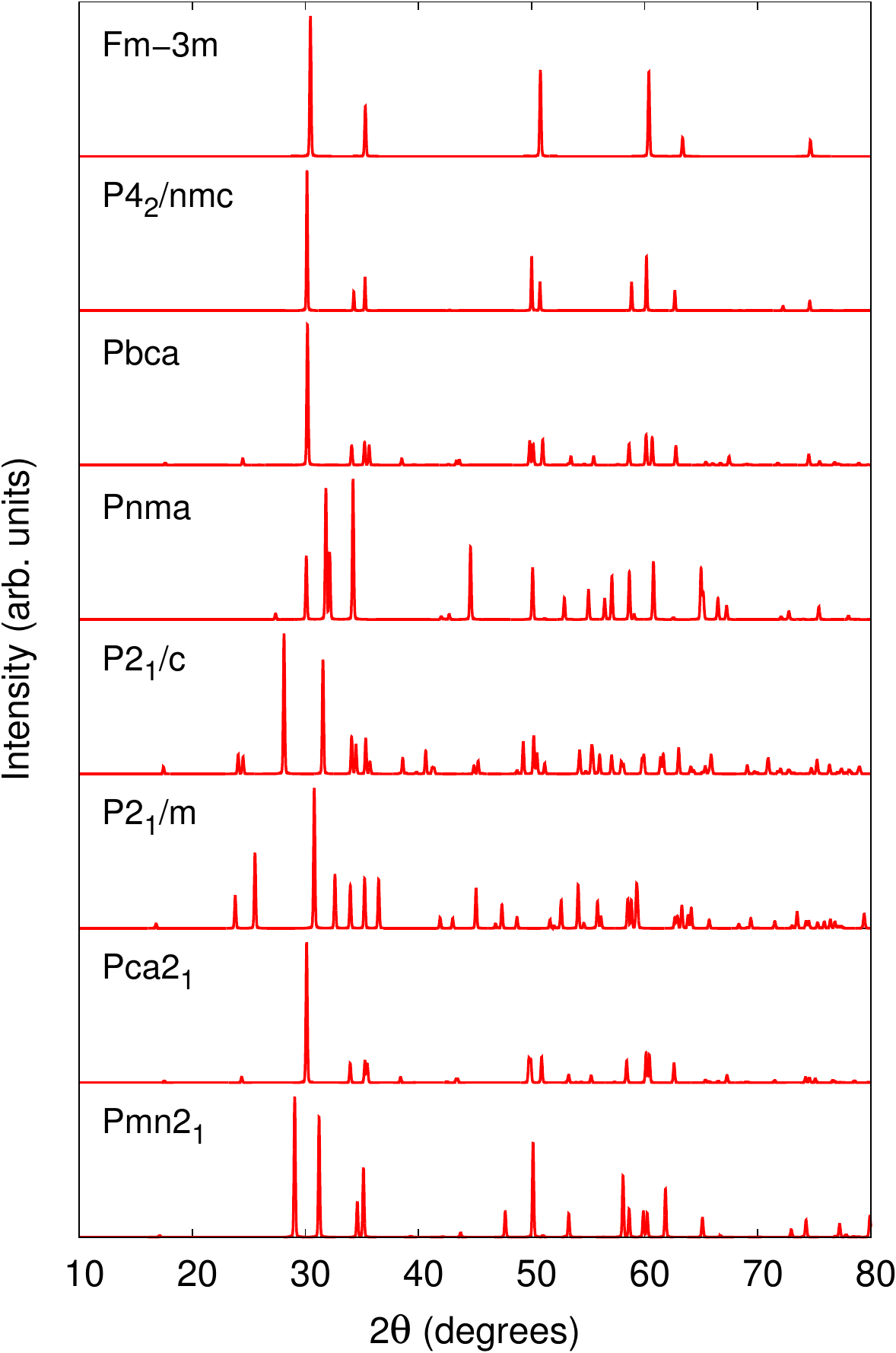}
\caption{Simulated x-ray diffraction patterns of the low-energy structures of hafnia examined in this work. The Cu K$\alpha$ (wavelength $\lambda=1.54$\AA) is used for calculations.}
\end{center}
\end{figure*}

\begin{figure*}[t]
 \begin{center}
\includegraphics[width=15cm]{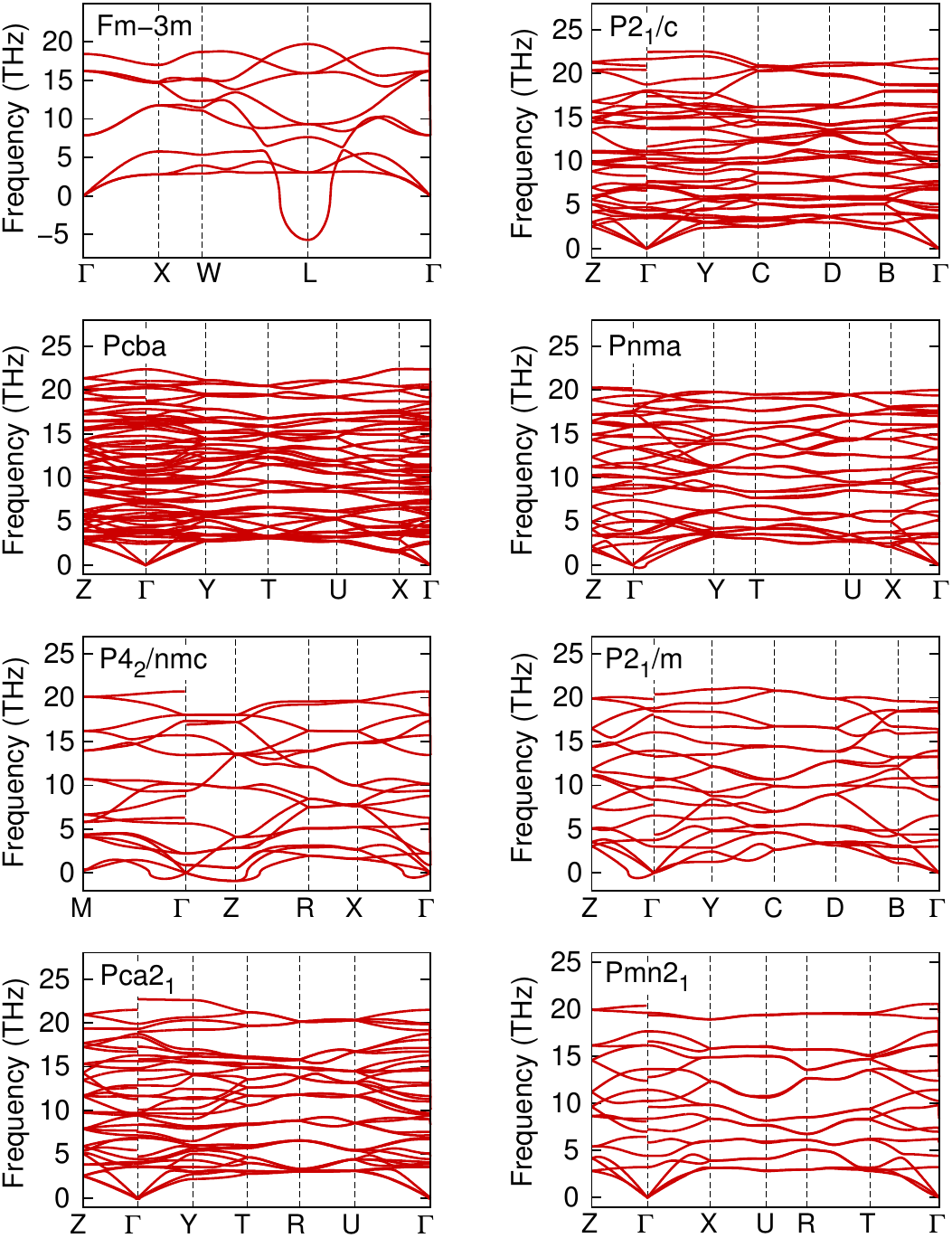}
\caption{Phonon band structures at $P=0$ GPa of eight low-energy structures of hafnia, plotted along the high-symmetry paths. For convenience, bands with imaginary frequencies are shown as those with {\it negative} frequencies.}
\end{center}
\end{figure*}